\newcommand{\version}{November 17, 2023}
\documentclass[letterpaper,oneside,11pt,pdftex]{article}
\pdfoutput=1
\usepackage[bindingoffset=0.0cm,textheight=22.60cm,hdivide={*,16.25cm,*}, vdivide={*,22.60cm,*}]{geometry}
\usepackage{amsmath,amsfonts,amssymb}
\usepackage{mathtools}
\usepackage[pdftex]{graphicx}
\usepackage[margin=1cm,font=small]{caption}
\usepackage[T1]{fontenc}
\usepackage[utf8]{inputenc}
\usepackage{fouriernc}

\usepackage[numbers,square,comma,sort&compress]{natbib}
\usepackage{import}

\usepackage[pdftex,hyperref,svgnames]{xcolor}
\usepackage[pdftex,bookmarksnumbered=true,breaklinks=true,%
colorlinks=true,linktocpage=true,linkcolor=MediumBlue,citecolor=ForestGreen,urlcolor=DarkRed]{hyperref}

\makeatletter
\AtBeginDocument{
  \hypersetup{
    pdfkeywords = {\keywords},
    pdftitle = {\@title},
    pdfauthor = {\@author}
  }
}
\makeatother

\makeatletter

 \@addtoreset{equation}{section}
 \makeatother

\newcommand{\ct}{c_{\textrm{T}}}
\newcommand{\cl}{c_{\textrm{L}}}

\renewcommand{\Im}{\mathrm{Im}}
\renewcommand{\Re}{\mathrm{Re}}

\title{\texorpdfstring{\begin{flushright}
        {\small LA-UR-23-25095}
       \end{flushright}\vspace{2em}}{}%
       Comparing theoretical predictions of radiation-free velocities of edge dislocations to molecular dynamics simulations}

\author{Daniel N. Blaschke\texorpdfstring{\footnotemark[2]~}{}, Ta Duong\texorpdfstring{\footnotemark[3]~}{}, Michael J. Demkowicz\texorpdfstring{\footnotemark[3]}{}}

\date{\version}

\newcommand{\keywords}{dislocations in crystals, dislocation mobility, crystal plasticity, transonic motion}

\graphicspath{{./}{./figures/}}

\begin{document}

 \maketitle

 \thispagestyle{empty}
 \begin{center}
 \renewcommand{\thefootnote}{\fnsymbol{footnote}}
 \vspace{-0.3cm}
\footnotemark[2] Los Alamos National Laboratory, Los Alamos, NM, 87545, USA\\
\footnotemark[3] Department of Materials Science and Engineering, Texas A\&M University, \\College Station, TX, 77843, USA
 \\[0.5cm]
 \ttfamily{E-mail: dblaschke@lanl.gov, ddinhta@tamu.edu, demkowicz@tamu.edu}
 \end{center}

\begin{abstract}
Transonic defect motion is of interest for high strain-rate plastic deformation as well as for crack propagation.
Ever since Eshelby's 1949 prediction in the isotropic limit of a 'radiation-free' transonic velocity $v_\text{RF}=\sqrt{2}\ct$, where shock waves are absent, there has been speculation about the significance of radiation-free velocities (if they truly exist) for defect mobility.
Here, we argue that they do not play any significant role in dislocation dynamics in metals, based on comparing theoretical predictions of radiation-free velocities for transonic edge dislocations with molecular dynamics simulations for two face-centered cubic (FCC) metals: Ag, where theory predicts radiation-free states, and  Cu, where it does not.
\end{abstract}

\tableofcontents

\section{Introduction and background}
\label{sec:intro}

At very high rates, plastic deformation is governed by high speed dislocations.
Dislocation mobility in this regime, however, is poorly understood  \cite{Hansen:2013,Luscher:2016,Alshits:1992,Blaschke:2019Bpap,Blaschke:2018anis,Blaschke:2019a}
and a key unresolved question is whether dislocations can reach transonic and supersonic speeds under sufficiently high stress; see also \cite{Weertman:1980,Gurrutxaga:2020} for reviews of high speed dislocation dynamics.
A suite of molecular dynamics (MD) simulations have indicated in recent years that such speeds are indeed possible \cite{Olmsted:2005,Marian:2006,Daphalapurkar:2014,Tsuzuki:2008,Tsuzuki:2009,Oren:2017,Ruestes:2015,Gumbsch:1999,Li:2002,Jin:2008,Peng:2019,Blaschke:2020MD,Dang:2022Mg}.
Experiments cannot yet track dislocations in metals in real time at these high speeds, but, with continued advances in time resolved in situ methods \cite{Wehrenberg:2017,Dresselhaus:2021,Katagiri:2023}, one can hope to indirectly determine the presence of supersonic dislocations and perhaps estimate the fraction and velocity of these dislocations in the near future.

An early cause for skepticism about the possibility of transonic or supersonic dislocation motion is that dislocation theory predicts divergences in self energy and stress at certain limiting velocities \cite{Blaschke:2021vcrit,Teutonico:1961, Teutonico:1962, Barnett:1973b} (at least while the core width is neglected \cite{Markenscoff:2008,Markenscoff:2009,Huang:2009}).
In the isotropic limit, these velocities coincide with the transverse and longitudinal sound speeds \cite{Eshelby:1949,Weertman:1980}.
In general, a dislocation moving at transonic or supersonic speed is expected to radiate sound waves because of the involved shock fronts.
However, in the isotropic limit Eshelby showed that within the theory of linear elasticity there exists a single transonic velocity for a gliding edge dislocation where this is not the case \cite{Eshelby:1949}.
The stress field of a moving edge dislocation in an isotropic solid in the continuum limit contains terms depending on the transverse and longitudinal sound speeds $\ct$ and $\cl$.
At a gliding velocity of exactly $v=\sqrt{2}\ct$, the term depending on $\ct$ is identically zero and hence the dislocation field is predicted to be 'radiation-free'.
Note that because the term in the dislocation field leading to shock waves is proportional to $\left(1-\frac{v^2}{2\ct^2}\right)$ \cite{Weertman:1967}, the radiation of elastic waves in the transonic regime tends to zero in a smooth fashion as  $v\to v_\text{RF} =\sqrt{2}\ct$, i.e. radiation energy is low in the vicinity of $v_\text{RF}$.
Much speculation about the relevance of this radiation-free state (if it truly exists) for both dislocation motion and crack propagation has appeared in the literature over the years, see \cite{Earmme:1974,Weertman:1980,Rosakis:1999,Rosakis:2001,Gumbsch:1999,Gao:1999,Gurrutxaga:2020,Cui:2022,Duong:2023a} and references therein.

Gao et al. \cite{Gao:1999} generalized Eshelby's calculation of radiation-free velocities to anisotropic crystals for both dislocation motion and crack propagation.
A closed form analytic solution was only found for systems exhibiting orthotropic symmetry (of which the isotropic limit is a special case), where the governing differential equations greatly simplify and only one transonic regime is present.
The basal and prismatic slip systems of HCP crystals, for example, exhibit orthotropic symmetry for gliding edge dislocations.
In more general settings, there exist three limiting velocities $c_3<c_2<c_1$ and hence two transonic regimes.
In the first (i.e. the lower) transonic regime, Gao et al. have shown formally that there always exists a radiation-free state for crack propagation.
For gliding dislocations, however, motion is restricted to the glide plane, and the radiation-free state requires a direction of dislocation motion which need not lie within the glide plane \cite{Gao:1999}.
In the second transonic regime, the existence of a radiation-free velocity depends crucially on the symmetry properties of the crystal, in particular the slip plane (in the case of dislocations).
The general (i.e., arbitrary symmetry) solutions discussed by Gao et al. in Ref. \cite{Gao:1999} are purely formal.

To our knowledge, no previous attempts have been made to determine from the very formal solutions given in Ref. \cite{Gao:1999} which (if any) anisotropic cubic crystals exhibit radiation-free velocities for gliding edge dislocations or to compare such calculations to dislocation mobilities in FCC or BCC metals obtained from MD simulations \cite{Gumbsch:1999,Olmsted:2005,Marian:2006,Tsuzuki:2008,Oren:2017,Peng:2019,Duong:2023a}.
This motivates us to search for radiation-free velocities in cubic and HCP crystals using numerical methods and to study their relevance for dislocation glide in the transonic regimes.
Results for selected metals are presented and a subset compared with our own MD simulations on two face-centered cubic (FCC) metals: copper (Cu) and silver (Ag).
Our analytical calculations predict a narrow range of radiation-free velocities for Ag and no radiation-free states for Cu.
Nevertheless, the dislocation mobilities of both metals revealed by MD simulations show no qualitative differences.
In particular, there is no evidence for a reduced drag force on the dislocation in the vicinity of the calculated radiation free velocity in Ag.
We conclude that the radiation-free velocities predicted in idealized theoretical settings (compact core, continuum, linear elasticity) do not occur in more realistic models (extended core, discrete lattice, non-linear elastic response).

\section{Finding radiation-free velocities from theory}
\label{sec:radiationfreetheory}

\begin{table}[ht]
\centering
\begin{tabular}{l|r|rrr||rrr|cc}
& $\rho\quad$&  $C_{11}\ \ $ & $C_{12}\ \ $ &  $C_{44}\ \ $ &  $c_3\ \ $ &  $c_2\ \ $ &  $c_1\ \ $ & $v_\text{RF}$ (1st) & $v_\text{RF}$ (2nd) \\
& [g/ccm] &  [GPa] & [GPa] &  [GPa] &  [km/s] &  [km/s] & [km/s] & [km/s] & [km/s] \\
\hline\hline
Ag &           10.50 &          123.99 &           93.67 &           46.12 &          1.20 &          2.10 &          3.84 
& $\sim2.09$ & none \\
 &           10.47 &          124.23 & 93.87 & 46.42  &          1.20 &          2.11 &          3.85 
& 2.09--2.10 & none \\
Al &            2.70 &          106.75 &           60.41 &           28.34 &          2.93 &          3.24 &          6.44 
& 2.93--3.10 & none\\
Au &           19.30 &          192.44 &          162.98 &           42.00 &          0.87 &          1.48 &          3.37 
& 1.39--1.47 & none \\
Cu &            8.96 &          168.30 &          121.20 &           75.70 &          1.62 &          2.91 &          4.96 
& none & none \\
 &            8.94 &          169.90 &          122.60 &           76.20 &          1.63 &          2.92 &          4.99 
& none & none \\
Ni &            8.90 &          248.10 &          154.90 &          124.20 &          2.29 &          3.74 &          6.05 
& none & none \\\hline
Fe &            7.87 &          226.00 &          140.00 &          116.00 &          2.75 &          2.93 &          6.41 
& none & 5.43\\
Mo &           10.20 &          463.70 &          157.80 &          109.20 &          3.61 &          3.68 &          6.30 
& 3.66--3.68 & 4.70\\
Nb &            8.57 &          246.50 &          134.50 &           28.73 &          2.15 &          2.34 &          4.95 
& 2.15--2.18 & 2.78\\
Ta &           16.40 &          260.20 &          154.40 &           82.55 &          1.93 &          1.96 &          4.28 
& none & 3.16 \\
W &           19.30 &          522.39 &          204.37 &          160.58 &          2.88 &          2.88 &          5.21
& none & 4.08
\\\hline
\end{tabular}
\caption{Material density, second order elastic constants and limiting velocities for edge dislocations in selected FCC (first five metals) and BCC (last five metals) cubic metals.
In the case of BCC metals, the limiting velocities are for edge dislocations in the \{110\} slip planes.
Most of the data shown here are taken from Ref. \cite{CRCHandbook}, except for typos in $C_{12}$ of Cu and $C_{44}$ of W which were corrected using the original references \cite{Epstein:1965} and \cite{Lowrie:1967}.
The second set of Ag values are determined from the Williams et al. EAM potential \cite{Williams:2006}.
The second set of Cu values are determined from the Mishin et al. EAM potential \cite{Mishin:2001}.
The limiting velocities were computed using PyDislocDyn \cite{pydislocdyn}; see \cite{Blaschke:2021vcrit} for a review on how to compute limiting velocities in general.
Our numerical implementation used to derive the radiation-free velocities in the first and second transonic regimes shown in this table as described in this section, has been included in PyDislocDyn \cite{pydislocdyn}.
In the case of almost-isotropic tungsten, it is easy to check that $c_3\approx c_2\approx v_\text{RF}/\sqrt{2}$ (i.e. $c_3=2875.03$ m/s, $c_2=2875.06$ m/s, and $v_\text{RF}/\sqrt{2}=2883.83$ m/s lies in the second transonic regime).
We do not find any radiation-free velocities for edge dislocations in the \{112\} nor the \{123\} slip planes of the BCC metals.}
\label{tab:vRFresultsgeneral}
\end{table}

\begin{table}[ht]
\centering
\begin{tabular}{l|r|r|r|r|r|r}
&  $\rho$ [g/ccm] &  $C_{11}$ [GPa] &  $C_{12}$ [GPa] &  $C_{13}$ [GPa] &  $C_{33}$ [GPa] &  $C_{44}$ [GPa]  \\\hline
Cd &           8.690 &          114.50 &           39.50 &          39.900 &           50.85 &           19.85  \\
Mg &           1.740 &           59.50 &           26.12 &          21.805 &           61.55 &           16.35  \\
Ti &           4.506 &          162.40 &           92.00 &          69.000 &          180.70 &           46.70 \\
Zn &           7.134 &          163.68 &           36.40 &          53.000 &           63.47 &           38.79  \\
Zr &           6.520 &          143.40 &           72.80 &          65.300 &          164.80 &           32.00 
\\\hline\hline
& $c_3$\,[km/s] &  \!$c_2$\,[km/s] &  \!$c_1$\,[km/s] & $v_\text{RF}$\,(1st)\,[km/s]
& $v_\text{RF}$\,(2nd)\,[km/s]
\\\hline
Cd, basal     &         1.51 &         (2.08) &         3.63 &      &       2.32 &\\
Mg, basal     &         3.07 &          (3.10) &         5.85 &       &        4.69 &\\
Ti, basal     &         (2.79) &         3.22 &          6.0 &          &     4.67 &\\
Zn, basal     &         2.33 &         (2.99) &         4.79 &         &     3.02 &\\
Zr, basal     &         2.22 &         (2.33) &         4.69 &         &     3.59 &\\
Cd, prismatic &         (1.51) &         2.08 &         3.63 &        &      2.94 
& $v_\text{RF}=\sqrt{2}c_2$ \\
Mg, prismatic &         (3.07) &          3.10 &         5.85 &          &    4.38 
& $v_\text{RF}=\sqrt{2}c_2$\\
Ti, prismatic &         2.79 &         (3.22) &          6.0 &           &   3.95 
& $v_\text{RF}=\sqrt{2}c_3$\\
Zn, prismatic &         (2.33) &         2.99 &         4.79 &         &     4.22 
&  $v_\text{RF}=\sqrt{2}c_2$\\
Zr, prismatic &         (2.22) &         2.33 &         4.69 &          &    3.29 
&  $v_\text{RF}=\sqrt{2}c_2$\\
Cd, pyramidal &         1.51 &         2.08 &         3.63 &           1.53--2.08 &  none & \\
Mg, pyramidal &         3.07 &          3.1 &         5.85 &                 none & none & \\
Ti, pyramidal &         2.79 &         3.22 &          6.0 &            2.8--2.95 & none & \\
Zn, pyramidal &         2.33 &         2.99 &         4.79 &           2.33--2.99 & none & \\
Zr, pyramidal &         2.22 &         2.33 &         4.69 &                 none & none &\\\hline
\end{tabular}
\caption{All data shown in this table are taken from Ref. \cite{CRCHandbook}.
The limiting velocities were computed using PyDislocDyn \cite{pydislocdyn}; see \cite{Blaschke:2021vcrit} for a review on how to compute limiting velocities in general.
The radiation-free velocities for the basal and prismatic slip systems can be calculated analytically from Eq. \eqref{eq:vRFresult}, as these slip systems are orthotropic which implies a complete decoupling of edge and screw components in the differential equations.
For this reason, only one of the two lower limiting velocities is ``seen'' by an edge dislocation, whereas the other one (indicated by brackets in the table above) would be the limiting velocity of a (hypothetical) screw dislocation with the same line orientation.
Therefore there is only one transonic region for edge dislocations in the basal and prismatic slip systems.
Our numerical implementation used to derive the radiation-free velocities shown for the pyramidal slip systems in this table as described in this section, has been included in PyDislocDyn \cite{pydislocdyn}.}
\label{tab:vRFresultshcp}
\end{table}


We start by briefly outlining the derivation of Gao et al. \cite{Gao:1999} before discussing our numerical implementation and results for various metals.
Restricting the discussion to steady state motion greatly simplifies the differential equations governing dislocation motion, because the time variable $t$ can be entirely eliminated from the dislocation displacement field $u_i(x,y,z,t)$ by a translation of the form $x'=x-vt$, where $x$ is the direction of dislocation motion at velocity $v$.
Choosing coordinates where $z$ is aligned with the dislocation line and $y$ is aligned with the slip plane normal, the following ansatz can be made for a straight dislocation of infinite length:
\begin{equation}
u_i = \Re\left(A_i f(x-vt+py)\right)
\,,
\end{equation}
where $A_i$ and $p$ are eigenvectors and eigenvalues within the framework developed by Stroh and others \cite{Stroh:1962,Bacon:1980}.
The displacement field $u_i$ is then substituted into the equations of motion
\begin{equation}
C'_{ijkl}\frac{\partial^2 u_k}{\partial x_j \partial x_l} = \rho \frac{\partial^2 u_i}{\partial t^2}
\,,
\end{equation}
where $\rho$ denotes the material density and $x_i=(x,y,z)$ so that we may at times use indices 1,2,3 and $x,y,z$ interchangeably in what follows.
Note that the $C'_{ijkl} = U_{ii'}U_{jj'}U_{kk'}U_{ll'}C_{i'j'k'l'}$ represent the tensor of second order elastic constants \emph{after rotation} into our present coordinates which have been aligned with the dislocation and slip plane instead of with the crystal coordinates; $U_{ij}$ denotes the rotation matrix.

The determinant of the resulting $3\times3$ coefficient matrix must vanish for non-trivial solutions of $A_k$.
In general, there exist three special dislocation velocities, denoted here as $c_3<c_2<c_1$, whose values depend on the gliding direction of the dislocation, where the solution $u_i$ becomes singular; see \cite{Blaschke:2021vcrit} for a review.
(In the isotropic limit, all directions are equivalent, $c_3=c_2$ equals the transverse sound speed $\ct$, and $c_1$ becomes the longitudinal sound speed $\cl$.)

Thus, there are four velocity ranges that lead to very different behaviors of the differential equations.
In particular, depending on the velocity $v$, the six eigenvalues $p$ appear as either
\begin{itemize}
\itemsep=0pt
\item three complex conjugate pairs in the subsonic regime ($v<c_3$),
\item or two complex conjugate pairs and two real values in the first transonic regime ($c_3\le v<c_2$),
\item or one complex conjugate pair and four real values in the second transonic regime ($c_2\le v<c_1$),
\item or six real values in the supersonic regime ($v\ge c_1$).
\end{itemize}
Gao et al. \cite{Gao:1999} order the eigenvalues $p_{\pm\alpha}$ ($\alpha=1,2,3$) such that $p_{\pm\alpha}$ becomes real when $v\ge c_\alpha$.
We are presently interested in the two transonic regimes.
Introducing the stress eigenvector $L_{\alpha,i}=C'_{i2kl} + p_\alpha C'_{i2k2}A_{\alpha,k}$, Gao et al. derive conditions the $L_{\alpha,i}$ must fulfill for a solution to be free of shock waves and hence radiation-free.
In particular, using only the subset of $L_{\alpha,i}$ for complex valued $p_\alpha$ (i.e. $\alpha=1,2$ in the first and $\alpha=1$ in the second transonic regime), it must be possible to fulfill the boundary conditions.
For an edge dislocation gliding in the $x$-direction, these are
\begin{align}
\lim_{y\to0^\pm}u_x(x-vt,y) &= \mp\frac{b}{2}\,,\qquad\forall (x-vt)>0
\,,\nonumber\\
\lim\limits_{y\to0}\sigma_{yy} &=0
\,,\label{eq:bc_edge}
\end{align}
where $b$ is the Burgers vector length, $u_x=u_1$, and $\sigma_{yy}=\sigma_{22}$.
In the second transonic regime, these translate to the conditions
\begin{align}
L_{1,2} = 0\,, \qquad\qquad
\Re\left[i a_1 L_{1,1}\right]=0 = \Re\left[i a_1 L_{1,3}\right]
\,,\label{eq:bc_second}
\end{align}
where $a_1$ is an arbitrary complex constant because the eigenvector $A_{1,i}$ is determined only up to a complex constant.
Eliminating the undetermined complex constant $a_1$, the last two conditions imply $\Re\left[L_{1,1}\right]\Im\left[L_{1,3}\right]=\Re\left[L_{1,3}\right]\Im\left[L_{1,1}\right]$.

In the first transonic regime, the conditions are
\begin{align}
\left(L_{1,2} + a_2L_{2,2}\right) = 0\,, \qquad\qquad
\Re\left[ia_1\left(L_{1,1}+a_2L_{2,1}\right)\right]=0 = \Re\left[ia_1\left(L_{1,3}+a_2L_{2,3}\right)\right]
\,,\label{eq:bc_first}
\end{align}
with two arbitrary complex constants $a_1$ and $a_2$.
Constant $a_2$ is determined from the first boundary condition, and the latter two conditions above imply
\[\Re\left[L_{1,1}+a_2L_{2,1}\right] \Im\left[L_{1,3}+a_2L_{2,3}\right]=\Re\left[L_{1,3}+a_2L_{2,3}\right]\Im\left[L_{1,1}+a_2L_{2,1}\right].\]
Though the $L_{\alpha,i}$ cannot in general be determined analytically (except for special cases discussed below), we can solve the relevant eigensystems numerically for any given velocity and subsequently check the boundary conditions we just discussed.
In the first transonic regime, one either finds a range of velocities that are radiation-free, or none at all, whereas the second transonic regime exhibits only one (if any) radiation-free velocities.
Our python implementation of the numerical search for radiation-free velocities is published as open source code within PyDislocDyn \cite{pydislocdyn} which is developed by one of us (DNB).

As with defect motion in general, the existence of radiation-free velocities in the first and second transonic regime depends strongly on the crystal and slip plane geometry.
The special case of orthotropic symmetry means the in-plane and out-of-plane crack deformation decouple from each other leaving only one transonic regime for crack propagation (see \cite{Gao:1999}).
In the context of dislocations, orthotropic symmetry allows the edge and screw components of $u_i$ to decouple such that only a $2\times2$ sub-matrix needs to be considered.
In this reduced system, only two limiting velocities and hence only one transonic regime are present for edge dislocations.
Therefore, the first two of conditions \eqref{eq:bc_second} apply (since $i=1,2$).
The radiation-free velocity can be determined analytically in this case to be
\begin{equation}
v_\text{RF} = \sqrt{\frac{C'_{11}C'_{22}-C_{12}'^2}{\rho\left(C'_{12}+C'_{22}\right)}}
\,, \label{eq:vRFresult}
\end{equation}
where we used Voigt notation for the elastic constants.

In the isotropic limit, the tensor $C'_{ijkl}=C_{ijkl}$ is invariant under rotations and since in this case $C_{22}=C_{11}$ and $C_{44}=\left(C_{11}-C_{12}\right)/2$, one recovers Eshelby's result of $v_\text{RF} = \sqrt{{2C_{44}}/{\rho}}=\sqrt{2}\ct$.

A similar result can be analytically derived for edge dislocations in prismatic slip systems of hcp crystals as indicated in the last column of Table \ref{tab:vRFresultshcp} above, which is reflective of the fact that the basal plane is isotropic with respect to its elastic constants and hence the radiation-free velocity is $\sqrt{2}$ times the limiting velocity:
For edge dislocations on prismatic slip planes, the rotated tensor of elastic constants aligned with the edge dislocation happens to coincide with the one aligned with the crystal coordinates so that $C'_{ijkl}=C_{ijkl}$.
Using the review article of Ref. \cite{Blaschke:2021vcrit} it is easy to work out that the analytical result for the lowest limiting velocity is $\sqrt{C_{66}/\rho}$ in this case.
Furthermore, since $C_{22}=C_{11}$ and $C_{66}=(C_{11}-C_{12})/2$ it is easy to see that \eqref{eq:vRFresult} reduces to $\sqrt{2 C_{66}/\rho}$.

Tables \ref{tab:vRFresultsgeneral} and \ref{tab:vRFresultshcp} show our results for a number of metals regarding radiation-free velocities.

\section{MD simulations}
\label{sec:MD}

\begin{figure}[!ht]
\centering
\includegraphics[trim=6cm 1.8cm 4.5cm 2.2cm,clip,width=0.75\textwidth]{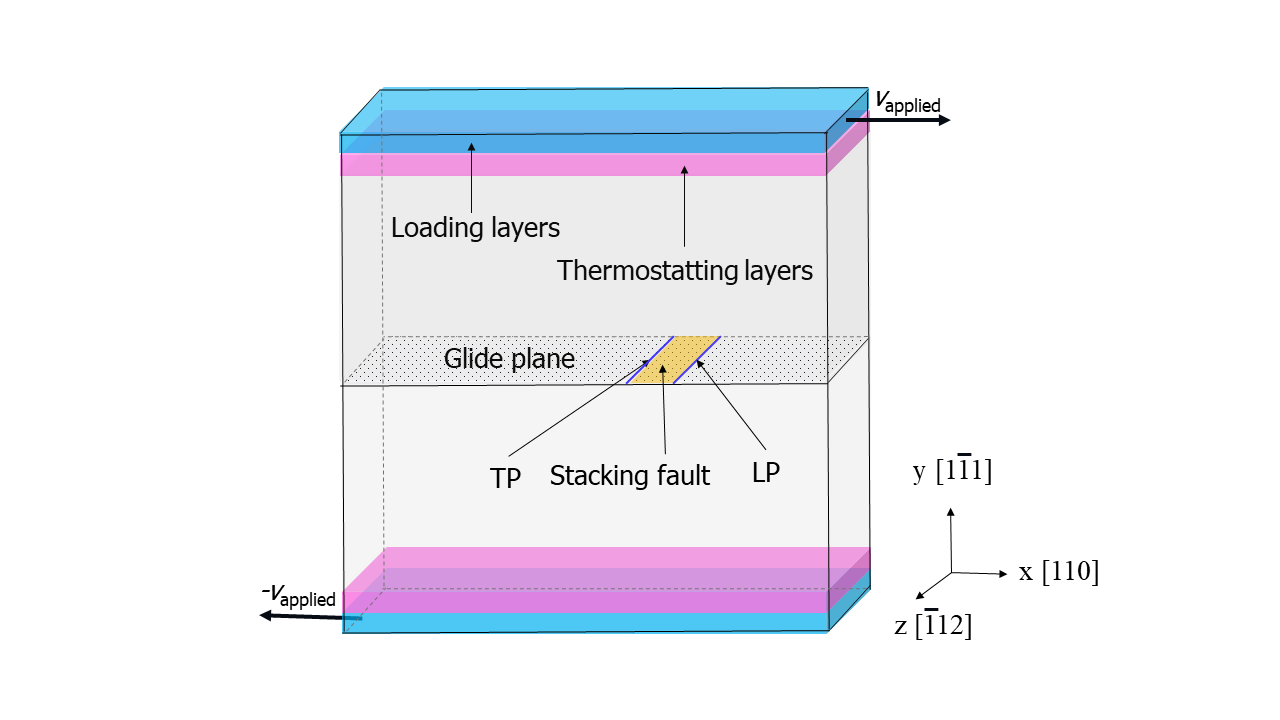}
\caption{Schematic of our atomistic models.}
\label{fig:schematic1}
\end{figure}

\noindent
To assess the effect of predicted radiation-free states, we perform molecular dynamics (MD) simulations of dislocation glide in copper (Cu) and silver (Ag) using the LAMMPS code \cite{Thompson:2022}.
As shown in Table \ref{tab:vRFresultsgeneral}, edge dislocation motion in Ag is predicted to have a radiation-free state in the first transonic regime while in Cu it is predicted to have no radiation-free states.
Therefore, comparing the behavior of these two materials allows us to assess whether the theoretically predicted radiation-free states have the expected effect on dislocation motion.
Interatomic forces are represented using the embedded atom method (EAM) \cite{Baskes:1984} potential for Cu by Mishin et al. \cite{Mishin:2001} and for Ag by Williams et al. \cite{Williams:2006}.
Both of these potentials are reportedly in good agreement with experiments \cite{Mishin:2001,Williams:2006} in terms of elastic constants, which are crucial for computing different characteristic velocities \cite{Blaschke:2021vcrit}, and in terms of stacking fault energies, which are important for modeling dislocations dissociated into Shockley partials \cite{Hull:2011}.
Visualizations are performed using Ovito \cite{Stukowski:2009}.

Figure \ref{fig:schematic1} shows a schematic of our atomistic models, which consist of a single crystal of either Cu or Ag with one edge dislocation inside.
Crystal orientations are chosen so that the $x$-axis is the dislocation propagation direction, the $y$-axis is normal to the glide plane, and the $z$-axis is parallel to the dislocation line.
Thus, the $x$-, $y$-, and $z$-axes align with the $[110]$, $[1\bar{1} 1]$, and $[\bar{1} 12]$ Miller index directions, respectively.
Periodic boundary conditions (PBCs) are applied in $x$- and $z$-directions.
In the $y$-direction, the model terminates with free surfaces.
The dimensions are $30\times25\times4$ nm$^3$ for the Cu model and $110\times55\times5$ nm$^3$ for the Ag model.
A larger Ag model is used to accommodate the larger separation between Shockley partials in this material.
The equilibrium distances between partials are 69\r{A} for Ag and 34\r{A} for Cu.
These separations agree with the experimental observations of Cockayne et al. \cite{Cockayne:1971}, where the stacking fault widths between partials are 70--90 \r{A} for Ag and 35--45 \r{A} for Cu.

\begin{figure}[!ht]
\centering
\includegraphics[trim=0 0.1cm 0 0.6cm,clip, width=0.65\textwidth]{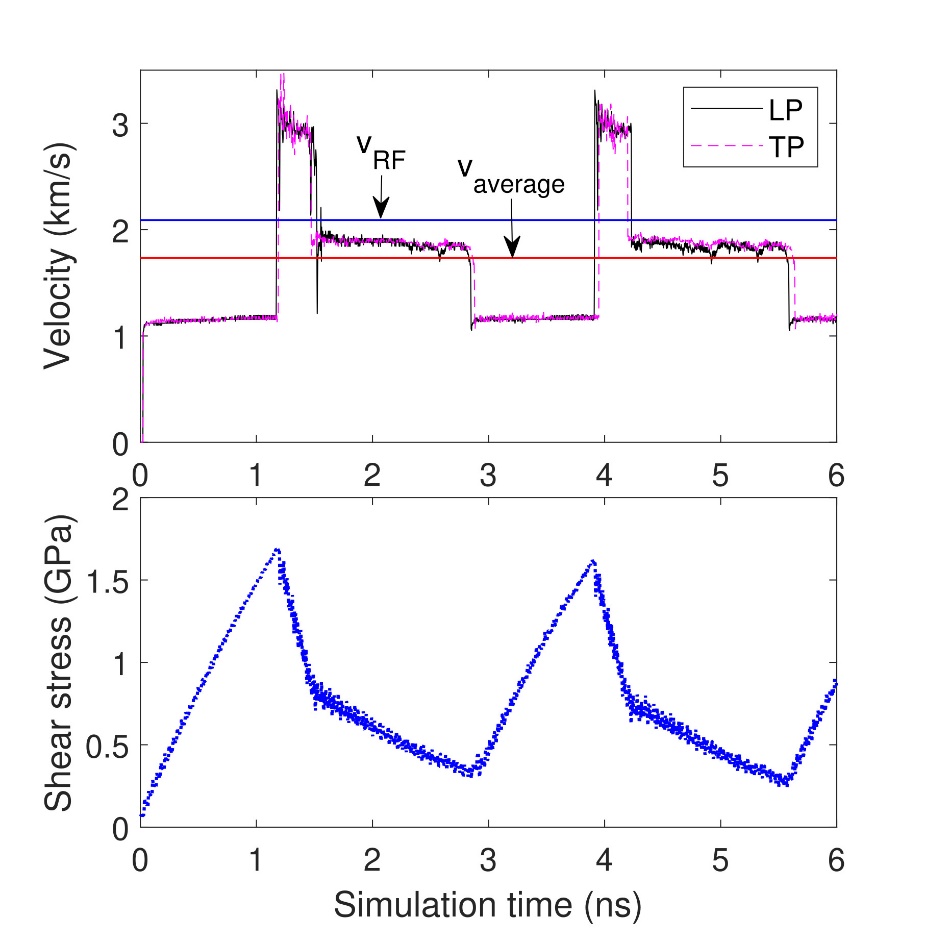}
\caption{Instantaneous dislocation velocity and surface traction for an edge dislocation in Ag at $v_\text{average} = 1.73$ km/s.}
\label{fig:silverA}
\end{figure}

Dislocations are made to move in the $x$-direction by displacing loading layers on each free surface at equal and opposite velocities, $v_\text{applied}$, parallel to the dislocation glide plane in the direction of the Burgers vector.
The constraint of velocity is applied to the center of mass of loading layers, not to individual atoms.
Tractions on the surfaces are computed from atomic forces in the loading layers divided by the surface area.
All simulations are initiated at 10K.
Thermostatting layers are located next to the loading layers to dissipate the heat generated from dislocation motion.
Thermostatting is performed by rescaling the temperature to 10K every 0.1ps.
A detailed description of our loading method is available in Ref. \cite{Duong:2023a}.  

According to Orowan's relation \cite{Argon:2007}, the average dislocation velocity can be computed as
\begin{equation}
v_\text{average} = \frac{2L_x}{b} v_\text{applied}
\,,
\end{equation}
where $L_x$ is the length of the model in the $x$-direction and $b$ is the length of the Burgers vector.
To investigate dislocations near $v_\text{RF}$ in Ag, we choose $v_\text{applied}$ so that the average velocity is in the first transonic regime: $c_3<v_\text{average} <c_2$.
Figure \ref{fig:silverA} shows the instantaneous velocity $v_i$ of the leading partial (LP), the trailing partial (TP) and the surface traction as a function of time for $v_\text{average} = 1.73$ km/s.
Similar to previous simulations carried out in Cu \cite{Duong:2023a}, the dislocation does not move with a uniform velocity.
Rather, its instantaneous velocity varies cyclically, with each cycle exhibiting abrupt jumps between three distinct branches:
the lowest branch with $v_i  < c_3$, a middle one with 1.9 km/s $< v_i < c_2$, and an upper one with 2.9 km/s $< v_i <$ 3.2 km/s.
The predicted radiation free velocity is near the upper bound of the middle branch.

\begin{figure}[!ht]
\centering
\includegraphics[trim=0.cm 0.5cm 0cm 0.4cm,clip, width=\textwidth]{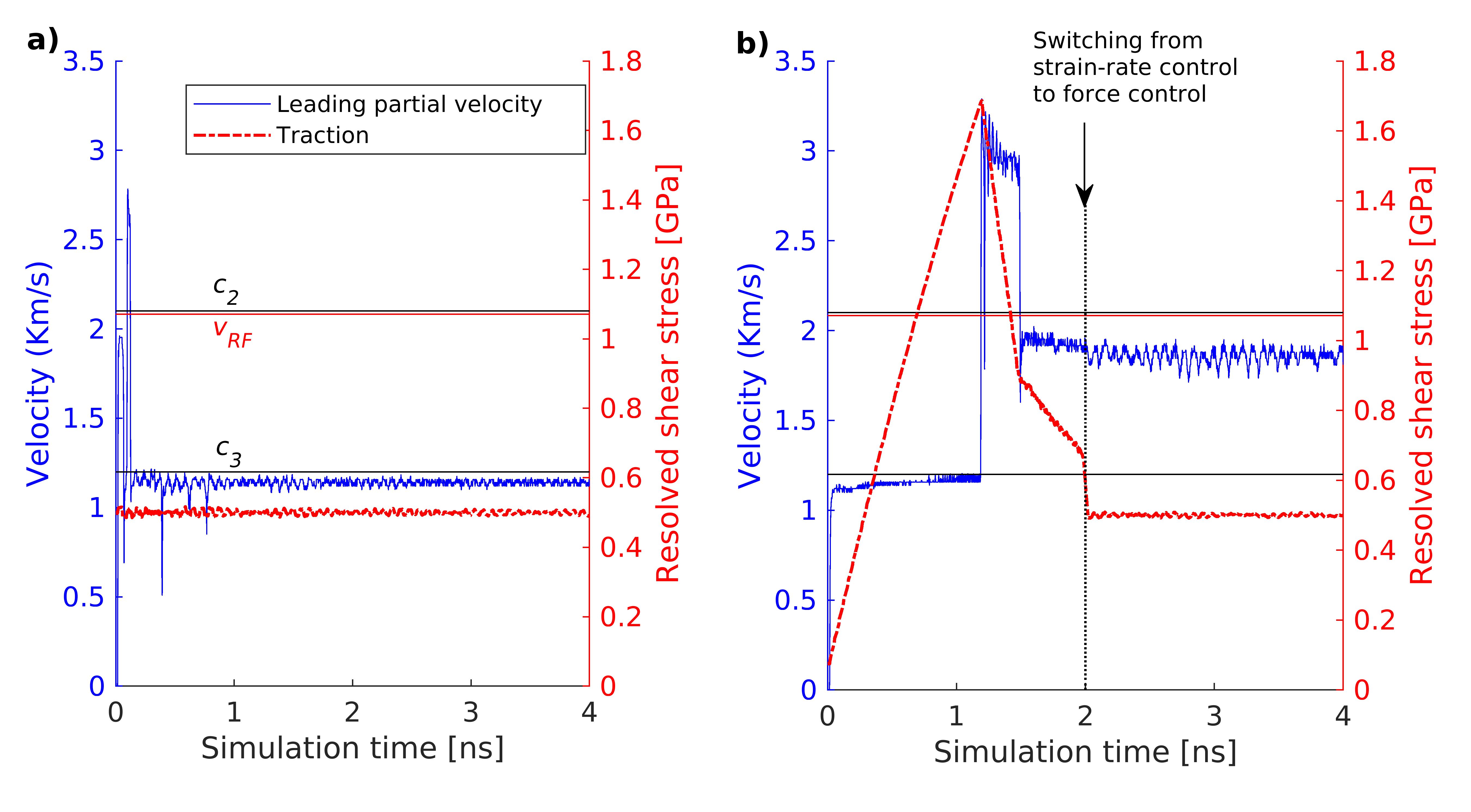}
\caption{We show steady state edge dislocations in Ag using the stress driven simulation method (a) as well as a hybrid method (b) where the dislocation is accelerated into the transonic regime using the strain rate driven method before switching to the stress driven method in order to maintain a steady state transonic edge dislocation.}
\label{fig:silversteady}
\end{figure}

Since the derivation from theory of the radiation free velocity assumed a steady-state gliding dislocation, whereas our strain rate driven simulations (Fig. \ref{fig:silverA}) exhibit only piece-wise near-constant velocity (whereas stresses vary continuously), we conducted also additional stress driven MD simulations for silver.
In particular, a constant shearing force was applied to the top and bottom layers of the MD simulation.
With this method, we were only able to access steady-state gliding velocities and stresses in the subsonic and supersonic regimes.
In order to achieve steady state transonic edge dislocations, we employed a hybrid method, where the dislocation was accelerated into the transonic regime with the strain rate driven method outlined above.
Then, we switched off the driving velocity and turned on a constant driving stress in our simulations to achieve a steady state transonic edge dislocation, see Fig. \ref{fig:silversteady}.
The simulation was repeated for multiple values of driving stress.

\begin{figure}[!ht]
\centering
\includegraphics[trim=0cm 0.2cm 0.cm 0.9cm,clip, width=0.85\textwidth]{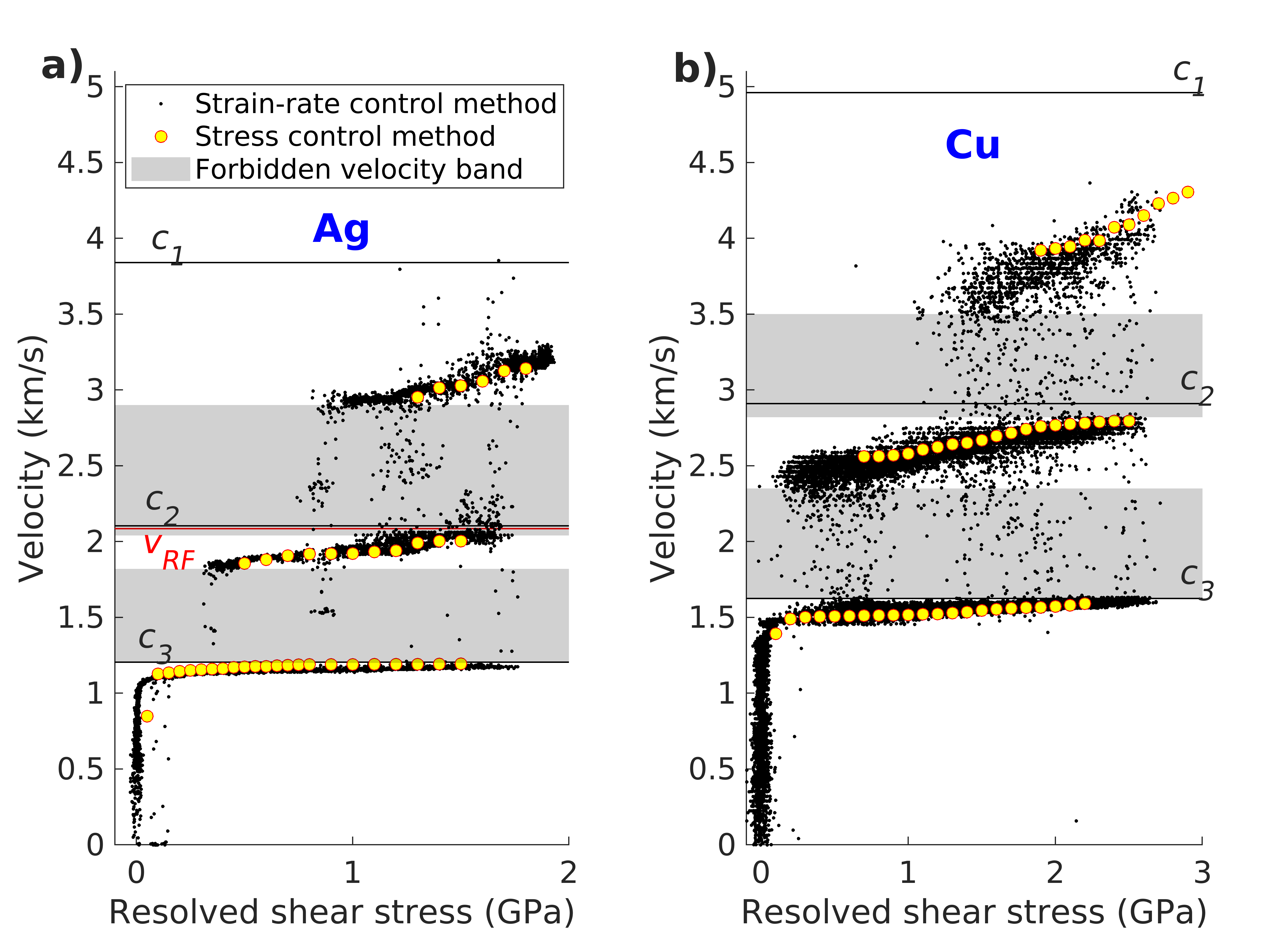}
\caption{The mobility relation of an edge dislocation in a) Ag and b) Cu \cite{Duong:2023a}.}
\label{fig:silverB}
\end{figure}

Mobility is the relationship between the resolved shear stress acting on a dislocation and the resulting uniform dislocation velocity.
It may be obtained from simulation results such as those shown in Fig. \ref{fig:silverA} by plotting the instantaneous velocity at each time against the shear stress at that same time (black data points).
From our additional stress driven simulations outlined above, we obtained steady state dislocation velocities versus constant driving stress (shown here in yellow) which agree with the strain rate driven values.
Figure \ref{fig:silverB}.a) shows the mobility relation of an edge dislocation in Ag assembled by plotting $v_i$ vs. resolved shear stress $\tau$ from 8 simulations with $v_\text{average}$ ranging from 0.1 km/s to 3.0 km/s.
This mobility relation is piece-wise continuous with three velocity branches separated by bands of forbidden velocities.
The isolated data points that appear within the forbidden bands correspond to rapidly accelerating dislocations, as they jump between different branches of the mobility relation.
Since these points represent a transient state of motion, they are not considered part of the mobility relation.

\begin{figure}[!ht]
	\centering
	\includegraphics[trim=0 0 0 0.3cm,clip,width=0.6\textwidth]{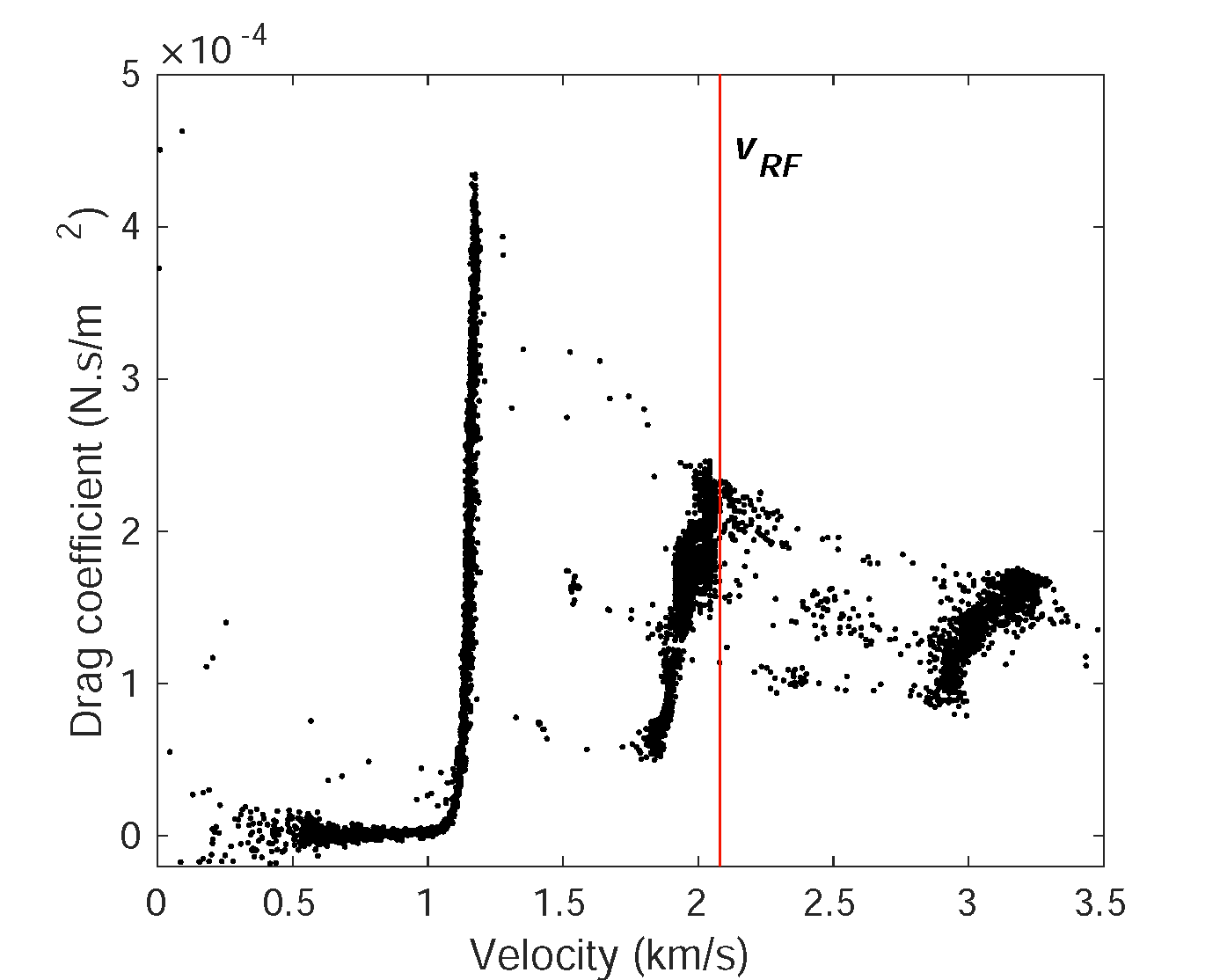}
	\caption{Drag coefficient as a function of velocity in Ag.}
	\label{fig:dragcoeffAg}
\end{figure}

Figure \ref{fig:silverB}.b) shows the mobility relation of Cu, described in Ref. \cite{Duong:2023a}.
Comparison of Fig. \ref{fig:silverB}.a) and \ref{fig:silverB}.b) shows that the mobility relations in Ag and Cu have the same qualitative form.
In particular, there is no apparent difference in the mobilities near the top of the middle branch, which is where $v_\text{RF}$ occurs in Ag.
Furthermore,
the presumed effect of a radiation-free state would be to reduce the drag force on a dislocation by decreasing the elastic energy dissipation from it.
Dislocation motion is governed by the quasi-Newtonian governing relation
\begin{equation}
m\dot{v} = \tau b - Bv
\,,
\end{equation}
where $m$ is dislocation mass per unit length, $\dot{v}$ is dislocation acceleration, and $B$ is the drag coefficient.
Assuming $\dot{v}=0$, the drag coefficient is
\begin{equation}
B = \frac{\tau b}{v}
\,.
\end{equation}
Based on the data in Fig. \ref{fig:silverB}.a), Figure \ref{fig:dragcoeffAg} shows that $B$ increases significantly in Ag as $v_i$ approaches $v_\text{RF}$.
Thus, contrary to the expectation from theory, the drag coefficient near $v_\text{RF}$ is not reduced.

\begin{figure}[!ht]
\centering
\includegraphics[trim=5cm 0 2cm 0,clip,width=0.85\textwidth]{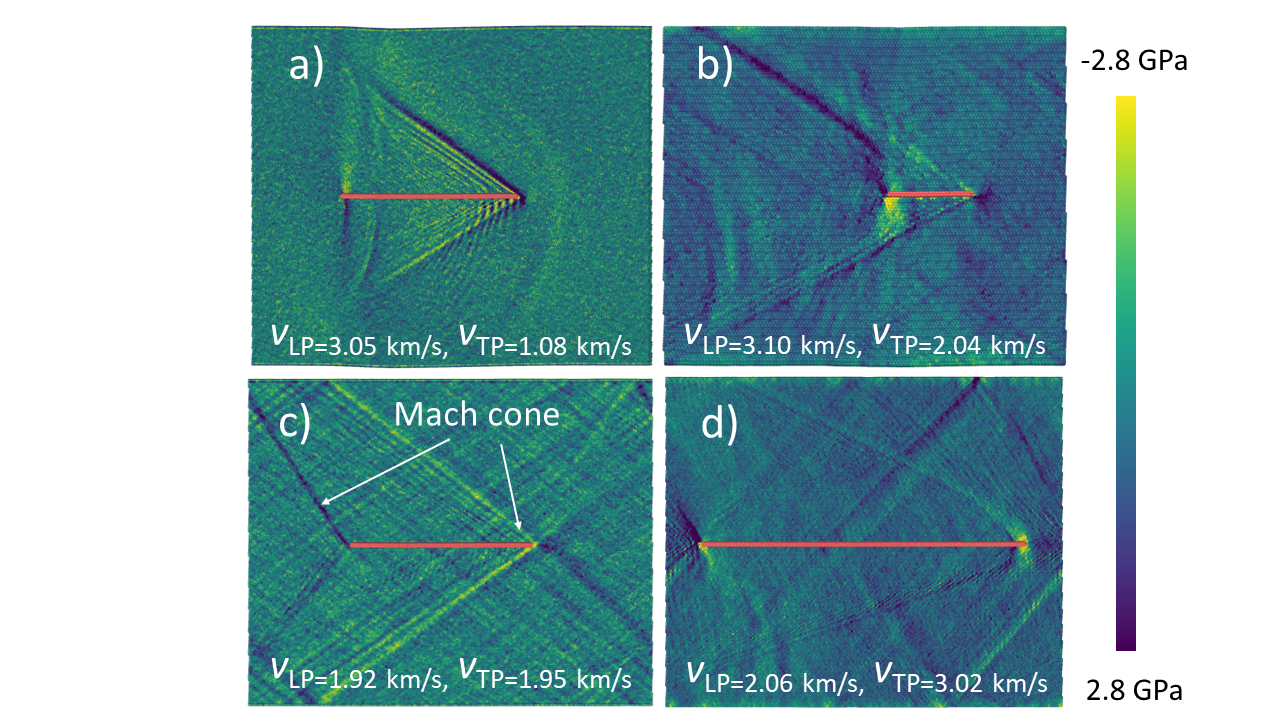}
\caption{Plots of $\sigma_{yy}$ allow us to identify shock waves generated by transonic dislocation in Ag.
Red atoms represent the stacking fault between LP and TP.
Dislocations move from left to right, so the LP is at the right-most termination of the stacking fault. Mach cones are easily seen in the local stress distribution, as illustrated in c).}
\label{fig:machcone}
\end{figure}

Linear elasticity theory predicts that dislocations traveling at $v_\text{RF}$ emit no shock waves \cite{Eshelby:1949,Gao:1999,Markenscoff:2001b}.
In atomistic simulations, such shock waves are easily seen as Mach cones \cite{Tsuzuki:2009}.
They have been reported in previous simulations in Cu \cite{Tsuzuki:2009} and W \cite{Li:2002}. 
Figure \ref{fig:machcone} visualizes Mach cones from the $\sigma_{yy}$ component of the instantaneous stress fields in our models at four different conditions:
\begin{itemize}
\itemsep=0pt
\item Fig. \ref{fig:machcone}.a): LP is transonic, TP subsonic, and the average of the LP and TP velocity is near $v_\text{RF}$.
In this case, only the LP emits a Mach cone.
\item	Fig. \ref{fig:machcone}.b): both LP and TP transonic, TP velocity near $v_\text{RF}$.
In this case, both partials emit Mach cones.
\item	Fig. \ref{fig:machcone}.c): both LP and TP transonic, both with velocity near $v_\text{RF}$.
Both emit Mach cones.
\item	Fig. \ref{fig:machcone}.d): both LP and TP transonic, LP velocity near $v_\text{RF}$.
Both emit Mach cones.
\end{itemize}

In summary, each partial emits a Mach cone whenever it is transonic, regardless of whether its velocity is near $v_\text{RF}$ where theory predicts no, or highly suppressed, Mach cones.
(As we pointed out earlier, theory predicts suppressed shock waves even in the vicinity of $v_\text{RF}$.)
Only the subsonic TP in Fig. \ref{fig:machcone}.a) does not emit a Mach cone.
We arrived at the same conclusions when examining Mach cones emitted by dislocations in Cu (not shown).
We conclude that realistic dislocation models (as opposed to the idealized ones of linear elasticity theory) do not cease to emit shock waves near $v_\text{RF}$.
Parenthetically, for Ag, the linear elastic model discussed in section \ref{sec:radiationfreetheory} predicts nearly identical $v_\text{RF}$ for both leading and trailing partials as it does for an undissociated edge dislocation.
Moreover, in Cu, it predicts no radiation free state for either the undissociated edge dislocation or the individuals Shockley partials into which it dissociates.

\begin{figure}[!ht]
\centering
\includegraphics[trim=0 0.8cm 0 0.9cm,clip,width=0.75\textwidth]{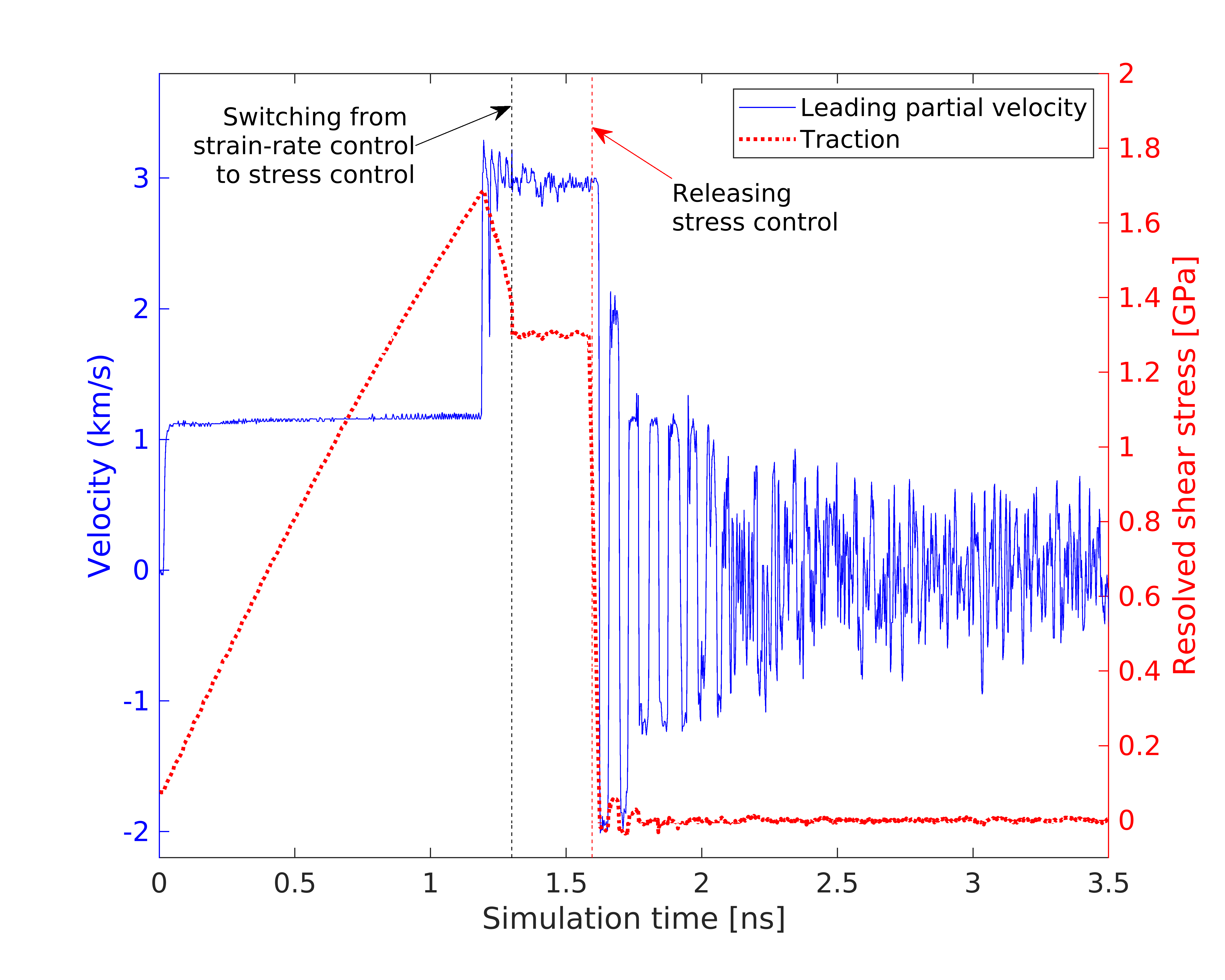}
\caption{Turning off the driving stress does not maintain any dislocation velocity, contrary to expectations at a `radiation free' velocity. (Only one exemplary simulation is shown here.)}
\label{fig:NoforceAg}
\end{figure}

As a final test, we carried out simulations where we (a) reduced the driving stress incrementally after achieving steady state dislocation glide and (b) turned off the driving stress after achieving steady state dislocation glide starting from both transonic branches.
When incrementally reducing the stress, the dislocation glide velocity decreases within a given branch before jumping to the next lower branch (i.e. from the second transonic to the first transonic and subsequently to the subsonic branch).
The lowest steady-state velocity in each branch correlated with the lowest driving stress in each branch indicating they are the velocities with the lowest drag.
Since neither of them is close to $v_\text{RF}$, we conclude that $v_\text{RF}$ is not the velocity of minimum drag.
When suddenly turning off the driving stress for a steady state dislocation in the second or first transonic branch, the dislocation velocity immediately drops and oscillates around zero, see Fig. \ref{fig:NoforceAg} showing the drop from the second transonic regime.
No steady-state velocity could be maintained at zero (or significantly reduced) driving stress, contrary to the theory prediction of a `radiation free' velocity, where it was conjectured \cite{Eshelby:1949} that a significant drop in drag (due to no elastic waves being emitted) would allow dislocation glide to continue at $v_\text{RF}$ with minimal driving stress.

At this point, we can speculate on why we do not see radiation-free states in MD simulations.
The theory derivation required a number of simplifying assumptions to be made and in principle any one (or several) of those could be the reason we find special radiation-free states which are not present in more realistic setups.
For example, it is known that divergences in the stress field at the limiting velocities can be removed when accounting for an extended dislocation core \cite{Markenscoff:2008,Pellegrini:2018}, so it stands to reason that the core could also prevent the shock wave producing part of the strain field solution from vanishing and hence remove any radiation-free velocity.
Furthermore, the theory of linear elasticity in the continuum limit is based on the assumptions that the displacements are not only much smaller than the lattice constants, but also slowly varying over the lattice, and hence the dimensionless gradients are small and slowly varying.
Also, non-linear effects such as the scattering of phonons on dislocations (an effect known as phonon wind) could potentially prevent radiation-free states even though the effect is small in our MD simulations since the number phonons is limited at the low temperature of 10K that we simulated.

\section{Discussion}

We presented an overview over theory predictions of so-called radiation-free velocities $v_\text{RF}$ for edge dislocations in a number of cubic and HCP metals.
Dislocation theory predicts the absence of shock waves at these special velocities which lie in the transonic regime and there has been a lot of speculation in the literature about the significance (if any) of these velocities for transonic dislocation motion.
To our knowledge, prior to this work, the numerical values of $v_\text{RF}$ had not been worked out for general anisotropic crystals, such as FCC and BCC metals, which have been of particular interest for studying transonic and supersonic dislocations from MD simulations.

Tables \ref{tab:vRFresultsgeneral} and \ref{tab:vRFresultshcp} list $v_\text{RF}$ for a selection of cubic and HCP metals.
Although hinted at in Ref. \cite{Gao:1999}, it does not seem to be widely known that in the first transonic regime there can be fairly large ranges of radiation-free velocities for gliding edge dislocations, depending on not only the crystal symmetry and slip systems, but also on the values of elastic constants.
For example, of the five FCC metals studied here, we found two (Cu and Ni) that did not exhibit any radiation-free velocity, whereas others (such as Al and Au) had radiation-free velocities predicted from theory over a range of values covering a significant portion of the first transonic regime.
Ag showed a very narrow range of radiation-free velocities in the first transonic regime in our theory calculations.
Similar results as in FCC metals are also found for pyramidal slip systems of HCP metals, whereas the basal and prismatic slip systems feature an orthotropic symmetry so that the analytic result, Eqn. \eqref{eq:vRFresult}, applies.
All BCC metals studied here exhibited the prediction of one radiation-free velocity in the second transonic regime, though only for the \{110\} slip systems, and only a subset of those metals exhibited the prediction of an additional range of radiation-free velocities in the first transonic regime.

Overall, however, our MD simulations found no evidence of anything special happening at the predicted radiation-free velocities, and also looking at MD results in the literature for other metals we find no evidence of radiation-free states:
\\
Olmsted et al. \cite{Olmsted:2005} find transonic edge dislocation motion in Al at a number of different velocities with no obvious correlation with any of our calculated radiation-free velocities.
Cu and Ni do not exhibit any radiation-free velocity according to our calculations presented in Table \ref{tab:vRFresultsgeneral}.
Transonic edge dislocations have been observed previously in MD simulations of both metals \cite{Olmsted:2005,Marian:2006,Tsuzuki:2008,Tsuzuki:2009,Daphalapurkar:2014,Oren:2017}.
We performed MD simulations on transonic edge dislocation motion in Cu and Ag.
In Ag, which is predicted to have a radiation-free state in the first transonic regime, we found transonic edge dislocations (see Fig. \ref{fig:silverB}), though their velocities did not correlate with the calculated (narrow) radiation-free velocity range (see Table \ref{tab:vRFresultsgeneral}).
The mobility function of edge dislocations in Cu, where no radiation-free states are predicted, was qualitatively similar to that in Ag, suggesting no effect of radiation-free states on edge dislocation motion.
In particular, there is no evidence of reduced dislocation drag near the radiation-free velocity in Ag.
Furthermore, we observed Mach cones near the predicted radiation free velocity in Ag (see Fig. \ref{fig:machcone}), where they should have been highly suppressed if $v_\text{RF}$ were truly `radiation free', and found no steady-state transonic gliding velocity which could be sustained with minimal driving stress.

Tungsten is elastically almost isotropic and transonic motion of edge dislocations has been studied in Ref. \cite{Gumbsch:1999}.
The radiation-free velocity is indeed predicted to be close to $\sqrt{2}$ times the shear wave speeds (which are almost identical, see Table \ref{tab:vRFresultsgeneral}), but only for the \{110\} slip planes.
The MD results of Ref. \cite{Gumbsch:1999} show transonic velocities ranging from 1.38$\ct$--1.5$\ct$ for edge dislocations, though only the \{112\} slip planes were studied in that work.
For tantalum, we find from theory a single radiation-free velocity for the \{110\} slip planes only.
Transonic edge dislocation motion was studied in Ref. \cite{Ruestes:2015}, where the authors find a range of transonic edge dislocation velocities.
In Mg, supersonic edge dislocations were reported in Ref. \cite{Dang:2022Mg} only for prismatic slip; no transonic dislocations were found and no dislocations moved at the radiation-free velocity which for prismatic slip is $\sqrt{2}c_2$ (i.e. significantly lower than the observed supersonic edge dislocation).

We conclude that non-linear effects such as phonon wind and dislocation core width are more important for dislocation dynamics than $v_\text{RF}$.
In fact, one might go as far as viewing the radiation-free velocities as a mathematical curiosity present only under (unrealistically) ideal conditions which include: vanishing core size, infinitely long and perfectly straight dislocations in the steady state limit, and the absence of non-linear effects.

\subsection*{Acknowledgements}
\noindent
DNB gratefully acknowledges support from the Materials project within the Physics and Engineering Models (PEM) Subprogram element of the Advanced Simulation and Computing (ASC) Program at Los Alamos National Laboratory (LANL). LANL, an affirmative action/equal opportunity employer, is operated by Triad National Security, LLC, for the National Nuclear Security Administration of the U.S. Department of Energy under contract 89233218NCA000001.

The atomistic modeling reported here was supported by the Department of Energy, Office of Science, Office of Basic Energy Sciences, under Award \#DE-SC0018892.

\bibliographystyle{utphys-custom}
\bibliography{dislocations}

\end{document}